\documentclass[aps,floats,floatfix,footnotes,prl]{revtex4} 
\usepackage{amstext} 
\usepackage{graphicx} 
\def\be{\begin{equation}} 
\def\ee{\end{equation}} 
\def\cR{{\cal R}}
\def\pf{{\rm pf}} 
\usepackage{amssymb}
 
\begin{document}

\title{Symmetry restoration in Hartree-Fock-Bogoliubov based theories
}
 
\author{G.F. Bertsch} 
\email{bertsch@uw.edu} 
\affiliation{Institute for Nuclear Theory and Dept. of Physics, Box 351560, University 
of Washington, Seattle, Washington 98915, USA} 
 
\author{L.M.~Robledo} 
\email{luis.robledo@uam.es} 
\affiliation{Departamento de F\'\i sica Te\'orica, M\'odulo 15, Universidad Aut\'onoma de 
Madrid, E-28049 Madrid, Spain} 
 
\begin{abstract} 
We present a pfaffian formula for projection and symmetry restoration
for wave functions of the general Bogoliubov form, including quasiparticle
excited states and linear combinations of them.  This solves a long-standing 
problem in calculating states
of good symmetry, arising from the sign ambiguity of the commonly used
determinant formula. A simple example is given of
projecting good particle number and angular momentum from a 
Bogoliubov wave function in the Fock space of a single $j$-shell.
\end{abstract} 
\maketitle

{\it Introduction.}  
The Bogoliubov transformation offers a powerful way to introduce correlations
into multi-fermion wave functions. The variational theory based on it, the 
Hartree-Fock-Bogoliubov (HFB) theory, has been very useful in nuclear physics.  
However, the variational wave functions need not respect symmetries of the 
Hamiltonian, hindering its use for spectroscopic purposes.  An
obvious fix is to project the wave functions onto eigenstates of the
conserved quantum numbers.  However, present methods to carry out the 
projection are beset with technical difficulties.  The purpose of this
letter is to present a projection formula that is applicable to general
Bogoliubov wave functions, including those for odd particle number. The
results are generalized for the evaluation of overlaps, as those required 
in configuration mixing theories based on HFB wave functions, commonly referred to as 
Generator Coordinate Method (GCM).

We first remind the reader that an operator ${\cal P}_K$
for projecting onto a symmetry group
representation $K$ is given by the integral
\be
{\cal P}_{Ki}  = {d_K\over \Omega_0} \int  d\Omega \,R^K_{ii}(\Omega) {\cal R}
(\Omega).
\ee
Here  $\cal R$ is an operator
of the symmetry group, $R^K_{ij}$  is a diagonal element of a matrix
representation of the group, $d_K$ is the 
dimension of the representation matrix $R^K$ and 
$\Omega_0 = \int d\Omega$ is the volume integral over the group.  The main 
conserved quantum numbers that we wish to restore in nuclear physics are
particle number $N$ and angular momentum $J$.  These are
both very familiar but for concreteness we note that 
particle number is associated with the gauge group $U(1)$ and the
group integral is $\int_0^{2\pi} d\phi$ where $\phi$ is the
gauge angle.  In the case of angular momentum, the integration is over
the Euler angles $\sin(\beta)d\alpha d\beta d\gamma$ and the representation
matrices are the Wigner $D$-functions.  
The probability of the component with quantum number $K$ in the state
$|w\rangle$ is given by the integral
\be
\langle K |w\rangle^2 = \langle w |{\cal P}_K | w \rangle = {d_K\over\Omega_0}\int  
d\Omega\, R^K_{ij} \langle w | {\cal R}|w\rangle.
\ee

In this letter we treat only the problem of calculating the overlaps;
for applications one also needs to calculate matrix elements
of physical operators. 
In the past, the computation of the overlap $\langle w| {\cal R}| w \rangle$
was carried out by the Onishi 
formula\cite{oy66} (see also \cite[Eq. E.49]{rs80}).
Unfortunately, the formula has square root sign ambiguity which makes
it useless for projection, except in special cases. 
Several suggestions have been made in the past for overcoming this
sign problem\cite{ne83,ha92,do98,ro09}.  In ref. \cite{ro09}, Robledo
proposed a promising new formula based on the pfaffian rather than the
determinant.  However, his formula requires the inverse of the
Bogoliubov transformation matrix $U$, and is thus not applicable 
to wave functions for which the $U$ matrix is singular.  This is the
case for all wave functions that have zero overlap with the vacuum.
In particular, the formula cannot be used directly for states of 
odd particle number.

Here we propose a pfaffian expression
which can be easily extended to odd-$N$ wave functions,
and indeed to states with more than one quasiparticle excitation.
To establish the notation, we write the effect of the symmetry operation
as
\be
\cR c^\dagger_i\cR^{-1} = \sum_j R_{ij} c^\dagger_j; \,\,\,\,\,\,\,
\cR c_i\cR^{-1} = \sum_j R^*_{ij} c_j 
\ee 
where $c^\dagger_i$ and $c_i$ are the usual Fock-space creation and
annihilation operator in some convenient basis.  Note that the matrix
$R$ depends on the specific details of the basis states and does not have 
to belong to an irreducible representation of the group.  The wave function
is characterized by the $U,V$ matrices of the Bogoliubov transformation.
Use is made of the Bloch-Messiah decomposition (see \cite{rs80} for details
and notation) that expresses
those matrices as the product of unitary matrices $D$ and $C$ and special
block diagonal matrices $\overline{\textrm U}$ and $\overline{\textrm V}$, namely 
$U=D\overline{\textrm U}C$ and $V=D^{*}\overline{\textrm V}C$ \cite[Eq. 7.8]{rs80}. 
The unitary $D$ transformation defines the "canonical" basis with 
creation and annihilation operators $a^{\dagger}$ and $a$.

We first consider the simpler case in which the wave function has a non-zero overlap with the
vacuum.  Then it can be expressed in the canonical basis as
\be
\label{wfn0}
|w\rangle = \prod^{n}_\alpha ( u_\alpha + v_\alpha a^\dagger_\alpha
a^\dagger_{\bar \alpha} ) |\rangle.
\ee
Here $n$ is the number of pairs in the wave function and the matrices
$U,V$ have dimension $(2n\times 2n)$.   
To specify the phase of the wave function, we may take all $u_\alpha$ 
positive definite.  
The overlap in this case is given by
\be
\label{normal}
\langle w| {\cal R}|w\rangle 
=
{(-1)^n\over \prod_\alpha^{n} (v_\alpha)^2} {\rm pf}
\left[ 
\begin{array}{cc}
  V^T        U & V^T       R^T V^* \\
-V^\dagger R V & U^\dagger     V^*
\end{array}
\right]
\ee
where $\pf(M)$ is the pfaffian of the matrix $M$.  
We outline an alternative derivation
below. Note that to use Eq. (\ref{normal})
the $U,V$ matrices in the canonical basis must be truncated to omit columns for which $v_\alpha=0$
(see also Ref. \cite{ro11}).
This simply means omitting the part of the Fock space that is not
occupied.

The generalization of Eq. (\ref{normal}) to deal with arbitrary overlaps is
straightforward but it requires to write the wave function $|w\rangle$ of
Eq. (\ref{wfn0}) as
\be 
|w\rangle= {\det C \over \prod_{\alpha=1}^n v_\alpha}\beta_1\beta_2\ldots\beta_{2n} |\rangle  
\label{wquasi}
\ee 
where the $\beta_\mu$ are Bogoliubov quasiparticle  annihilation operators with 
amplitudes $U$ and $V$.
Here an unnormalized wave function is obtained by the product of all the 
Bogoliubov-transformed annihilation operators acting
on the vacuum, and $\prod^n_{\alpha=1} v_\alpha^{-1}$ is the normalization 
factor. The phase $\det C$ is required for consistency of all the definitions.
The overlap is then given by
\be
\label{normalo}
\langle w| {\cal R}|w ' \rangle 
=
(-1)^n {\det C^{*} \det C' \over \prod_\alpha^{n} v_\alpha v'_\alpha} {\rm pf}
\left[ \begin{array}{cc}
 V^T          U & V^T        R^T V'^* \\
-V'^\dagger R V & U'^\dagger     V'^*
\end{array}
\right]
\ee
This formula is useful in dealing with configuration
mixing of symmetry restored HFB wave functions, as required in the implementation of
the most general version of the Generator Coordinate Method (GCM). The connection
between Eq. (\ref{normalo}) and Eq.  (7) of \cite{ro09} is not straightforward
and requires of some lengthy calculations, details are given in \cite{supp}.

Eq. (\ref{normalo})  may be extended to wave functions that are 
orthogonal to the vacuum by considering the more general canonical form
\be
\prod_q c^\dagger_q \prod^{n}_\alpha ( u_\alpha + v_\alpha a^\dagger_\alpha
a^\dagger_{\bar \alpha} ) |\rangle.
\ee
Again, if the canonical $U,V$ ($U',V'$) matrices are truncated to omit columns for
which $v_\alpha=0$ ($v'_\alpha=0$), Eq. (\ref{normalo}) is still applicable.  
In the case of 
odd-$N$ ground states, only a 
single additional operator is needed,
\be
|qw\rangle = c^\dagger_q |w\rangle = \sum_j q_j c^\dagger_j|w\rangle.
\ee  
We use the notation $\bf q$ for the row vector of the coefficients $q_i$ 
($q_{1,i}$ in matrix notation), and
${\bf 0}$ for the row vector of zeros.
Then the generalization of Eq. (\ref{normalo}) is
\be
\label{qp}
\langle qw| \cR | q'w'\rangle =
(-1)^n {\det C^{*} \det C' \over \prod_\alpha^{n} v_\alpha v'_\alpha} {\rm pf}
\left[ \begin{array}{cccc}
        V^TU     &      {\bf 0}^T                  &     V^T R^T {\bf q'}^T    &       V^T R^T V'^*  \\
      {\bf 0}    &          0                      &  {\bf q}^* R^T {\bf q'}^T &  {\bf q}^* R^T V'^* \\
  -{\bf q}' R V  & -{\bf q}' R {\bf q}^\dagger     &        0                  &       {\bf 0}       \\
-V'^\dagger R V  &  - V'^\dagger R{\bf q}^\dagger  &     {\bf 0}^T             &  U'^\dagger V'^*    \\
\end{array}\right].
\ee  
The shape of this matrix is $(2n+2+2n)\times(2n+2+2n)$.
To derive it and Eq. (\ref{normalo}), first
note that the expectation value of a product of single-fermion operators
$\alpha_i$ is given by the pfaffian of all possible contractions\cite{li68,ca73}
\be
\langle \alpha_1\alpha_2\ldots\alpha_k \rangle  = {\rm pf}\left(\,
S_{i, j} \,\right)
\ee
where $S_{i, j}$ is a skew-symmetric matrix with upper triangular elements 
$S_{i, j} = \langle |\alpha_i \alpha_j | \rangle$ ($i < j$)
The overlaps can be written in this form using Eq.  (\ref{wquasi}).
The overlap in Eq. (\ref{qp}) is derived by evaluating the contractions in the
operator product
\be
\langle \beta^\dagger_{2n}\beta^\dagger_{2n-1}\ldots\beta^\dagger_1 c_q 
 \tilde{c}^{\dagger}_{q'} 
\tilde{\beta}'_1\tilde{\beta}'_2\ldots\tilde{\beta}'_{2n}\rangle
\ee 
where $ \tilde{\beta}' = {\cal R} \beta' {\cal R}^{-1}$ etc. The matrices
entering the pfaffian of Eq. (\ref{qp}) are easily identified: 
\begin{eqnarray}
(V^{T}U)_{\mu \nu} & = & \langle | \beta^{\dagger}_{\mu} \beta^{\dagger}_{\nu} | \rangle \\
(V^{T} R^{T} {\bf q'}^T)_{\mu} & = & 
\langle | \beta^{\dagger}_{\mu} \tilde{c}^{\dagger}_{q'} | \rangle = 
\sum_{j} q'_{j} \langle | \beta^{\dagger}_{\mu} \tilde{c}^{\dagger}_{j} | \rangle \\
(V^{T}R^TV'^*)_{\mu \nu} & = & \langle | \beta^{\dagger}_{\mu} \tilde{\beta}'_{\nu} | \rangle \\
{\bf q}^* R^T {\bf q}'^T & = & \sum_{j j'} q^*_{j} q'_{j'} \langle | c_{j} \tilde{c}^{\dagger}_{j'} | \rangle \\
({\bf q}^* R^T V'^*)_\nu & = & \sum_{j}  q^*_{j} \langle | c_{j} \tilde{\beta}'_{\nu} | \rangle \\
(U'^{\dagger}V'^*)_{\mu \nu} & = & \langle | \tilde{\beta}'_{\mu} \tilde{\beta}'_{\nu} | \rangle.
\end{eqnarray}

The generalization of Eq.  (\ref{qp}) to multi-quasiparticle overlaps, with
$r$ annihilation operators $\bar{\beta}_{\mu_j}$ (Bogoliubov amplitudes 
$\bar{U}, \bar{V}$) to the left of $\cR$ and
$s$ creation operators $\bar{\beta}'_{\nu_j}$ to the right, is tedious
but straightforward
\be
\label{mqp}
\langle w| \bar{\beta}_{\mu_r} \cdots \bar{\beta}_{\mu_1} \cR \bar{\beta}'^\dagger_{\nu_1} \cdots \bar{\beta}'^\dagger_{\nu_s} | w'\rangle =
(-1)^n (-1)^{r(r-1)/2} {\det C^{*} \det C' \over \prod_\alpha^{n} v_\alpha^* v'_\alpha} {\rm pf}
\left[ \begin{array}{cccc}
        V^TU     &         V^T{\bf p}^\dagger          &     V^T R^T {\bf q'}^T        &       V^T R^T V'^*  \\
   -{\bf p}^* V  &     {\bf q}^*  {\bf p}^\dagger      &  {\bf q}^* R^T {\bf q'}^T     &  {\bf q}^* R^T V'^* \\
  -{\bf q}' R V  & -{\bf q}' R {\bf q}^\dagger         &     {\bf p}' {\bf q}'^T       &       {\bf p}' V'^* \\
-V'^\dagger R V  &  - V'^\dagger R{\bf q}^\dagger      &    -V'^\dagger {\bf p}'^T     &  U'^\dagger V'^*    \\
\end{array}\right].
\ee  
For this expression to make sense both $r$ and $s$ must have the same number parity. 
The objects ${\bf p}$ and ${\bf q}$ (${\bf p}'$ and ${\bf q}'$) are matrices of 
dimension $r\times 2n$ ($s\times 2n$) with
matrix elements $p_{\mu_j m}  = \bar{V}_{m \mu_j}$ and $q_{\mu_j m} = \bar{U}_{m \mu_j}$. 
If some of the $\bar{\beta}$ annihilation operators are replaced by creation ones $\bar{\beta}^\dagger$
the appropriate rows of ${\bf q}$ and ${\bf q}$ have to be redefined accordingly.
It is easy to check that Eq.  (\ref{mqp}) reduces to Eq.  (\ref{qp}) in the limit ${\bf p}=0$.
Apart from the fact that Eq.  (\ref{mqp}) includes the phase of the matrix element, this
expression has the advantage over more traditional approaches \cite{ba69} that
the combinatorial explosion in the evaluation of the left hand side of Eq. (\ref{mqp}), 
namely the fact that $(r+s)!!$ contractions have
to be considered if the multi-quasiparticle overlap is computed with the
standard Generalized Wick's theorem, is completely avoided (see \cite{pe07} for
another approach based on the finite temperature formalism).

{\it Example.}  As a proof of principle, we carry out the projection for 
an odd-$N$ wave function having a nontrivial structure with respect
to angular momentum and particle number.  We take the Fock space
as the 6-dimensional space of orbitals in a $j=5/2$ shell.  The creation
operators $c_m^\dagger$ are labeled by azimuthal angular momentum
$j_z=m$.  The wave function for the test is
\be
\label{wf}
|qw\rangle = c_{1/2}^\dagger\left(u + v c_{5/2}^\dagger c_{-5/2}^\dagger
\right) |\rangle
\ee
with $(u,v) = (0.8,0.6)$.  We project simultaneously
on particle number and angular momentum with the operator ${\cal P}_N
{\cal P}_{JJ_z}$.  We use a 4-point uniform mesh for integrating the 
gauge angle and a 5-point Gauss-Legendre
quadrature for integrating over the angular variable $\cos(\beta)$.
There is
no necessity to integrate over the other Euler angles because
the wave function Eq. (\ref{wf}) is an eigenstate of $J_z$.  
The results are 
shown  in Table I.  
The projected quantum numbers $N$ and
$J$ are given in the first two columns. The third column gives the
exact decomposition, and the fourth column the numerical projection.
One sees that there is complete agreement to the level of numerical
precision in the integrations.
\begin{table}[htb]
\caption{Test of number and angular momentum projection for the wave
function of Eq. (\ref{wf}), for which $J_z=1/2$.    
}\begin{tabular}{|cc|lc|}
\hline
 & & \multicolumn{2}{|c|}{$ \langle N J J_z | qw\rangle^2$}\\
$N$ & $J$ & analytic   & numerical \\
\hline
1 &  3/2  &  0          & 0.00000   \\
1 &  5/2  & $u^2 = 0.64$ &  0.64000   \\
3 &  1/2  &  0          & 0.00000  \\
3 &  3/2  &  $v^2/7\approx 0.05143$ & 0.05143  \\
3 &  5/2  &  $v^2/2=0.18 $ & 0.18000 \\
3 &  7/2  &  0   & 0.00000  \\
3 &  9/2  &  $5v^2/14\approx 0.12857$   & 0.12857 \\
\hline
\end{tabular}
\end{table}

{\it Discussion.} Besides the overlap function
$\langle w | {\cal R} |w\rangle$ we need the matrix elements
of various operators in the symmetry-restored states.  For the most
important operators they can be expressed as a single integral over
matrix elements of the type
$\langle w |{\cal O} {\cal R} |w\rangle$
where $\cal O$ is an operator such as the Hamiltonian.  
It is straightforward to calculate this operator matrix elements 
using the Balian-Brezin formula\cite{ba69} or the multi-quasiparticle
overlap of Eq.  (\ref{mqp}).  Unlike the formulation
in Ref. \cite{ro09},
our method here does not require one to construct quasiparticle
states explicitly.  Our procedure is easily extended to
multi-quasiparticle matrix elements with a final result that avoids
the combinatorial explosion that plagues other methods used to evaluate those
overlaps.
For an $k$-quasiparticle excitation, the pfaffian matrix is augmented
by $2k$ rows and columns.  A program that demonstrates the method for
the example in Table I is provided in the supplementary material.
After this work was posted on arXiv \cite{arXiv} we learned that a similar 
fully general formula for the overlap was obtained independently 
by Avez and Bender \cite{av11} and Oi and Mizusaki \cite{oi11}.

{\it Acknowledgment.}  GFB thanks H. Goutte and F. D\"onau for discussions and
correspondence, and A. Bulgac for comments on the manuscript.
This work was supported in part by the U.S. Department of Energy under Grant 
DE-FG02-00ER41132, and by the National Science 
Foundation under Grant PHY-0835543. 
The work of LMR was supported by MICINN (Spain) under  
grants Nos. FPA2009-08958, and FIS2009-07277, as well as by  
Consolider-Ingenio 2010 Programs CPAN CSD2007-00042 and MULTIDARK  
CSD2009-00064.  
 
\appendix*

\section{Supplementary material}

We want to prove the equivalence between the result of Eq. (7) 
in the letter for the special case ${\cal R}=1$
\be \label{Eq1}
\langle w| w'\rangle=(-1)^n
{\det (C^*) \det (C') 
\over 
\prod_\alpha^n v_{\alpha}v'_{\alpha} }
\pf \left(\begin{array}{cc}
V^{T}U & V^{T}V'^*\\
-V'^{\dagger} V & U'^{\dagger}V'^{*}
\end{array}\right).
\ee
and the overlap formula of Eq (7) in Ref. \cite{ro09}. In both cases 
$U,V$ are the $2n\times 2n$ Bogoliubov transformation matrices and $C$ 
the third (unitary) transformations of their Bloch-Messiah decomposition. 
The normalized wave function $|w\rangle$ is written as 
\be
|w\rangle = {\det (C) \over \prod_\alpha^{n} v_\alpha} |\tilde{w}\rangle 
\ee
with $|\tilde{w}\rangle = \beta_1 \ldots \beta_{2n} |\rangle$. Using the Bloch-Messiah
decomposition it is easy to show that $|\tilde{w}\rangle$ and the wave function 
$$|\phi \rangle = \exp \left(\frac{1}{2} \sum_{k k'} M_{k k'} c^{+}_{k} c^{+}_{k'}\right) |\rangle$$
of Eq (1) in \cite{ro09} are related by 
\be
|\tilde{w}\rangle = \pf (U^\dagger V^*) |\phi\rangle
\ee 
In terms of $|\tilde{w}\rangle$
\be \label{Eq2}
\langle \tilde{w}| \tilde{w}'\rangle=(-1)^n
\pf \left(\begin{array}{cc}
V^{T}U & V^{T}V'^*\\
-V'^{\dagger} V & U'^{\dagger}V'^{*}
\end{array}\right).
\ee
In Ref. \cite{ro09}, Eq (7) the following equation is derived for the overlap.
\be \label{target}
\langle \phi | \phi ' \rangle=(-1)^n \pf \left(\begin{array}{cc}
V'^*U'^{*\,-1} & -\mathbb{I} \\
\mathbb{I}& -V U^{-1}\\
\end{array}\right)=(-1)^n\pf \left(\begin{array}{cc}
M'& -\mathbb{I} \\
\mathbb{I}& -M^*\\
\end{array}\right)
\ee
where we write $ M = (VU^{-1})^*$ and $N$ in Eq (7) of  \cite{ro09} 
is twice the $n$ here ($N=2n$, and therefore $S_{N}=(-1)^{N(N+1)/2}=(-1)^{n}$).
To prove the equivalence of the two expressions, we make multiple use
of the pfaffian identity for the congruence transformation,
\be \label{PI1}
\pf (A^T B A) = \det (A) \,\, \pf (B).
\ee
We first write the matrix in Eq. (\ref{Eq2}) as
\be
\left(\begin{array}{cc}
V^{T}U & V^{T}V'^*\\
-V'^{\dagger}V & U'^{\dagger}V'^{*}
\end{array}\right)=\left(\begin{array}{cc}
V^{T} & 0\\
0 & V'^{\dagger}
\end{array}\right)\left(\begin{array}{cc}
UV^{-1} & \mathbb{I}\\
-\mathbb{I} & -U'^{*}V'^{*\,-1}
\end{array}\right)\left(\begin{array}{cc}
V & 0\\
0 & V'^{*}
\end{array}\right)
\ee 
Then using Eq. (\ref{PI1}) 
the overlap in Eq. (\ref{Eq2}) can be expressed as 
\be \label{Eq5}
\langle \tilde{w}| \tilde{w}'\rangle=
(-1)^n \det (V) \det (V'^*) \pf \left(\begin{array}{cc}
M^{*\,-1} & \mathbb{I}\\
-\mathbb{I} & -M'^{-1}
\end{array}\right).
\ee

We next use the congruence transformation to express the matrices in 
Eq. (\ref{Eq5}) and (\ref{target}) as
\begin{equation}
\left(\begin{array}{cc}
M^{*\,-1} & \mathbb{I}\\
-\mathbb{I} & -M'^{-1}
\end{array}\right)=\left(\begin{array}{cc}
\mathbb{I} & 0\\
-M^{*} & \mathbb{I}
\end{array}\right)\left(\begin{array}{cc}
M^{*\,-1} & 0\\
0 & -M'^{-1}+M^{*}
\end{array}\right)\left(\begin{array}{cc}
\mathbb{I} & M^{*}\\
0 & \mathbb{I}
\end{array}\right)\label{eq:trick}
\end{equation}
\begin{equation}
\left(\begin{array}{cc}
M' & -\mathbb{I}\\
\mathbb{I} & -M^{*}
\end{array}\right)=\left(\begin{array}{cc}
\mathbb{I} & 0\\
M'^{-1} & \mathbb{I}
\end{array}\right)\left(\begin{array}{cc}
M'  & 0\\
0 & -M^{*}+M'^{-1}
\end{array}\right)\left(\begin{array}{cc}
\mathbb{I} & -M'^{-1}\\
0 & \mathbb{I}
\end{array}\right)\label{eq:trick2}.
\end{equation}
The congruence transformation matrices have determinant equal to one,
so the pfaffians of left-hand side are are equal to those of the
transformed matrices, 
$\pf (M^{*\,-1})\pf (-M'^{-1}+M^{*})$ and
$\pf (M')\pf (M'^{-1}-M^{*})$,
respectively. With these results and the properties 
$\pf (A^{-1}) = (-1)^n / \pf (A)$
and $\pf (-A) = (-1)^n \pf (A)$ we get
\be \label{Eq5b}
\langle \tilde{w}| \tilde{w}'\rangle=
{ \det (V) \det (V'^*) \over \pf (M^*) } \pf (-M'^{-1}+M^{*})
\ee
and
\be \label{targetb}
\langle \phi | \phi ' \rangle=
 \pf (M')  \pf (-M'^{-1}+M^{*})
\ee
The equivalence between the two expressions 
Eqs. (\ref{Eq5b}) and (\ref{targetb}) 
is evidenced by taking into account that 
\be
{\det (V) \over \pf (M^*)} = \det (V^T) \pf (V^{T\, -1} U^T) = \pf (U^T V)
\ee 
which finishes the proof.

The generalization to the case where ${\cal R}$ is not the unity matrix,
proceeds along the same lines.


\end{document}